# On the correct mathematical proof of the polarization mode dispersion equation

Carlos L. Janer[1,*]

[1]Department of Electronic Engineering, Escuela Superior de Ingenieros,Universidad de Sevilla, Camino de los Descubrimientos s/n, Seville, Spain
*Corresponding author: janer@us.es

*Abstract*—The fundamental equation that describes polarization mode dispersion does not have a mathematically correct and convincing convincing proof. This problem stems from the fact that Poincare's sphere, where Stokes vectors are represented, is just a manifold (a representation space) devoid of metric. In this "space" orthogonal vectors are antiparallel and, therefore, it is hard to justify the use of a Euclidean metric. However, if one realizes that in Poincare's sphere only three-dimensional rotations are represented, a Euclidean pseudo-scalar product can be defined and a mathematically correct proof for the polarization mode dispersion can be given.
OCIS-Codes:060.0060.

Polarization mode dispersion is one of the physical processes that limit high bit rates in optical communication systems [1-4]. Single mode fibers support two modes of propagations that are orthogonally polarized and are, in principle, degenerate that is to say, they both travel with the same group velocity. However, random optical birefringence due to imperfection in the manufacturing process cause different group delays for different input polarization states [1-4]. This phenomenon is called polarization mode dispersion (PMD). It has traditionally been described by the following well known differential equation, where all vectors are defined in a "space" that contains Poincare's sphere (since we will be dealing with fully polarized light Stokes "vectors" only need to have three components) [1-4]:

$$\frac{\partial \vec{\tau}}{\partial z} = \vec{\beta}_{\omega} + \vec{\beta} \times \vec{\tau} \quad (1)$$

Where $\vec{\tau}$ represents the PMD vector, z the distance travelled along the fiber, $\vec{\beta}$ the fiber birefringence vector, the letter subscript ω indicates differentiation, and × stands for the cross product so that $\vec{\beta} \times \vec{\tau}$ represents the rotation of $\vec{\tau}$ due to the birefringence of the fiber.

The geometrical representation of monochromatic waves in Poincare's sphere has the following properties: each state of polarization (SOP) is related to a unique "pseudo-vector" $\hat{s}$ on the surface of this sphere. Orthogonal SOPs are 180° apart from one another [7]. The birefringence of fibers is represented by non-unitary vectors ($\vec{\beta}$) pointing in the same direction as the slow local eigenstate. The PMD vector is defined by another non-unitary vector ($\vec{\tau}$) pointing at the slow global eigenstate . In Fig. [1] an SOP denoted by $\hat{s}$ and the vectors $\vec{\beta}$ and $\vec{\tau}$ along with the angles defining $\hat{s}$ are represented.

It should be stressed the fact that, in opposition to Jones vectors which do belong to a two dimensional complex vector space, Stokes "vectors" do not form a vector space [7]. Moreover the group structure of the proper three-dimensional group of rotations, SO(3,R), is not isomorphic to the special unitary group SU(2). There is an homomorphism of SU(2) onto SO(3,R), not an isomorphism [6]. To check the fact that Stokes vectors do not belong to a vector space, one has only to add two states of polarization of equal length along the S1 and –S1 coordinates in Stokes "space" and see that a 45° linear SOP is not obtained.

Equation (1) has, most of the times, been mathematically "proven" as follows:

$$\frac{\partial \hat{s}}{\partial z} = \vec{\beta} \times \hat{s}$$
$$\frac{\partial \hat{s}}{\partial \omega} = \vec{\tau} \times \hat{s} \quad (2)$$

The first equation means that the SOP rotates around the local birefringence vector as the wave propagates along the fiber (see Fig.[1]). The second one states that, as the wave's angular velocity changes, the SOP rotates around the local PMD vector (see, also, Fig. [1]). Since both rotations must be compatible, there has to be relationship between $\vec{\tau}$ and $\vec{\beta}$ which is supposed to be equation (1). The traditional "proof" starts differentiating the first equation with respect to ω and the second with respect to z. The second order cross derivates are eliminated and, finally, the following identities are used [4]:

$$\vec{\beta} \times (\vec{\tau} \times \hat{s}) = \vec{\tau}(\vec{\beta} \cdot \hat{s}) - \hat{s}(\vec{\beta} \cdot \vec{\tau})$$
$$\vec{\tau} \times (\vec{\beta} \times \hat{s}) = \vec{\beta}(\vec{\tau} \cdot \hat{s}) - \hat{s}(\vec{\tau} \cdot \vec{\beta}) \quad (3)$$

However, this procedure is not satisfactory and mathematically sound. The reason underlying this fact is that Poincare's sphere, where SOPs are represented, is just a manifold which is, in principle, devoid of metric and therefore of scalar product. The polarization states (Stokes unitary three-dimensional pseudo-vectors), usually denoted as $\hat{s}$, do not even belong to a vector space since the addition of two SOPs is not, in general, a unitary vector and therefore does not belong to Poincare's sphere [7]. Moreover, if Poincare's sphere is embedded in a vector space, orthogonal SOPs are represented by anti-parallel Stokes vectors and, therefore, it would be paradoxical to define a Euclidean metric in this space. However the mathematical "proof" of (1) relies heavily on the following Euclidean cross product identity which is ill defined because dot products are not unambiguous operations (which metric should be used in this space and why?): $\vec{a}\times(\vec{b}\times\vec{c}) = \vec{b}(\vec{a}\cdot\vec{c}) - \vec{c}(\vec{a}\cdot\vec{b})$. This problem plagues most papers on PMD because dot products are used very often. This paper proves rigorously that a pseudo-Euclidean dot product can be safely used.

It will also be shown in this letter that the only mathematical assumption needed to prove (1) is that Poincare's sphere is just a parametric manifold embedded in a space where three-dimensional rotations are represented. The partial derivates of $\hat{s}$ with respect the three coordinate axes generate the three elementary rotations needed to represent any differential three-dimensional rotation. The specific expressions of these three generator $(\breve{L}_1, \breve{L}_2, \breve{L}_3)$ given in (5) are needed to induce a pseudo-scalar product in this metric-free parametrical space as it will be shortly proven.

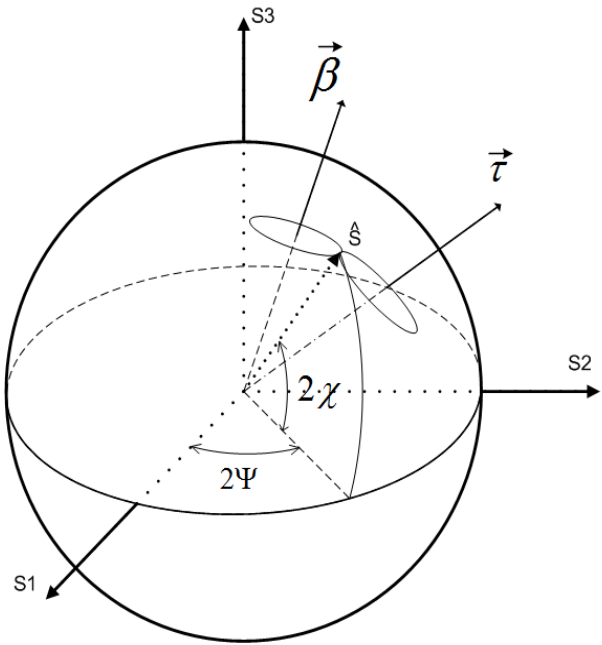

Fig. 1. Poincare's sphere. The angles ψ and χ represent the orientation of the principle axes in Jone's space and the ellipticity

respectively of the SOP $\hat{s}$. $\vec{\beta}$ is the birefringence vector and $\vec{\tau}$ represents the PMD vector.

To the best of my knowledge, there is just one proof [5] which does not make use of any metric or scalar product but, unfortunately, its validity is restricted to the case where $\vec{\beta}$ and $\vec{\beta}_\omega$ are parallel.

An approach like [5] will also be used in this letter to prove (1), that is, to take into account that the rotations in a three-dimensional space form a group, but taking one step further: to bear in mind that it is a very special group, a Lie group whose commutators' expressions are very well known.

It should be stressed the fact that the lack of a general mathematical proof of (1) could raise doubts about its validity and that these doubts should be either eliminated or confirmed. This letter proves that (1) has indeed a mathematically sound proof and, therefore, is a completely valid general equation. It also proves that a pseudo-Euclidean dot product can be safely used.

Let us begin giving the set of commutation relations of the three dimensional group of rotations and their expressions in an orthogonal frame of reference:

$$[\breve{L}_1,\breve{L}_2] := \breve{L}_1\cdot\breve{L}_2 - \breve{L}_2\cdot\breve{L}_1 = \breve{L}_3$$
$$[\breve{L}_2,\breve{L}_3] := \breve{L}_2\cdot\breve{L}_3 - \breve{L}_3\cdot\breve{L}_2 = \breve{L}_1 \quad (4)$$
$$[\breve{L}_3,\breve{L}_1] := \breve{L}_3\cdot\breve{L}_1 - \breve{L}_1\cdot\breve{L}_3 = \breve{L}_2$$

$$\breve{L}_1 = \begin{bmatrix} 0 & 0 & 0 \\ 0 & 0 & -1 \\ 0 & 1 & 0 \end{bmatrix}$$

$$\breve{L}_2 = \begin{bmatrix} 0 & 0 & 1 \\ 0 & 0 & 0 \\ -1 & 0 & 0 \end{bmatrix} \quad (5)$$

$$\breve{L}_3 = \begin{bmatrix} 0 & -1 & 0 \\ 1 & 0 & 0 \\ 0 & 0 & 0 \end{bmatrix}$$

Equations (2) may be stated in a very compact way:

$$[\breve{L}_i,\breve{L}_j] = \varepsilon_{ijk}\breve{L}_k \quad (6)$$

where $\varepsilon_{ijk}$ is a third order anti-symmetric tensor whose values are +1 if ijk form an even permutation, -1 if the permutation is odd and 0 otherwise. $[\breve{L}_i, \breve{L}_j]$ represents the commutator of the operators $\breve{L}_i$ and $\breve{L}_j$.

It should also be noted that:

$$Tr[\breve{L}_i \cdot \breve{L}_j] = -2\delta_{ij} \qquad (7)$$

where Tr stands for the trace of the matrix product.

In the following proof the first and final equations will be expressed both in their traditional way (infinitesimal rotations are defined by a cross product) and in their more correct way (infinitesimal rotations are defined by the action of the Lie group generators):

$$\frac{\partial \hat{s}}{\partial z} = \frac{\partial \hat{s}}{\partial x_1}\frac{\partial x_1}{\partial z} + \frac{\partial \hat{s}}{\partial x_2}\frac{\partial x_2}{\partial z} + \frac{\partial \hat{s}}{\partial x_3}\frac{\partial x_3}{\partial z} =$$
$$= \beta^1 \frac{\partial \hat{s}}{\partial x_1} + \beta^2 \frac{\partial \hat{s}}{\partial x_2} + \beta^3 \frac{\partial \hat{s}}{\partial x_3} = \qquad (8)$$
$$= \beta^1 \breve{L}_1 \hat{s} + \beta^2 \breve{L}_2 \hat{s} + \beta^3 \breve{L}_3 \hat{s} =$$
$$= \sum_1^3 \beta^i \breve{L}_i \hat{s} \equiv \beta^i \breve{L}_i s$$

$$\frac{\partial \hat{s}}{\partial w} = \frac{\partial \hat{s}}{\partial x_1}\frac{\partial x_1}{\partial \omega} + \frac{\partial \hat{s}}{\partial x_2}\frac{\partial x_2}{\partial \omega} + \frac{\partial \hat{s}}{\partial x_3}\frac{\partial x_3}{\partial \omega} =$$
$$= \tau^1 \frac{\partial \hat{s}}{\partial x_1} + \tau^2 \frac{\partial \hat{s}}{\partial x_2} + \tau^3 \frac{\partial \hat{s}}{\partial x_3} = \qquad (9)$$
$$= \tau^1 \breve{L}_1 \hat{s} + \tau^2 \breve{L}_2 \hat{s} + \tau^3 \breve{L}_3 \hat{s} =$$
$$= \sum_1^3 \tau^i \breve{L}_i \hat{s} \equiv \tau^i \breve{L}_i \hat{s}$$

where Einstein's index summation convention has been used and will be used, when necessary, from now on. It should be noted that, since in this space only rotations are represented, the partial derivatives with respect the coordinates represent infinitesimal rotations around the corresponding axis. If equation (8) is derived with respect ω and equation (9) with respect z the following expressions are obtained:

$$\frac{\partial^2 \hat{s}}{\partial \omega \partial z} = \beta^i_\omega \breve{L}_i \hat{s} + \beta^i \tau^j \breve{L}_i \breve{L}_j \hat{s}$$
$$\frac{\partial^2 \hat{s}}{\partial z \partial \omega} = \tau^i_z \breve{L}_i \hat{s} + \tau^i \beta^j \breve{L}_i \breve{L}_j \hat{s} \qquad (10)$$

In order to ensure the compatibility of the rotations expressed in (8) and (9), these two equations are subtracted and forced to be equal to zero:

$$0 = (\beta^i_\omega - \tau^i_z)\breve{L}_i \hat{s} + \beta^i \tau^j (\breve{L}_i \breve{L}_j - \breve{L}_j \breve{L}_i)\hat{s} =$$
$$= (\beta^i_\omega - \tau^i_z)\breve{L}_i \hat{s} + \beta^i \tau^j [\breve{L}_i, \breve{L}_j]\hat{s} = (\beta^k_\omega - \tau^k_z)\breve{L}_k \hat{s} \quad (11)$$
$$+ \beta^i \tau^j \varepsilon_{ijk} \breve{L}_k \hat{s} = (\beta^k_\omega - \tau^k_z + \beta^i \tau^j \varepsilon_{ijk})\breve{L}_k \hat{s}$$

Since (11) is valid for any polarization state $\hat{s}$ and $\breve{L}_k$ is a non-null operator, the following expression is deduced:

$$\beta^k_\omega - \tau^k_z + \beta^i \tau^j \varepsilon_{ijk} = \frac{\partial \vec{\beta}}{\partial \omega} - \frac{\partial \vec{\tau}}{\partial z} + \vec{\beta} \times \vec{\tau} = 0 \qquad (12)$$

which is exactly the equation that I was trying to prove.

There is, however, more information that can be obtained from (7) which is commonly derived from its vector formulation evaluating its dot product with $\vec{\tau}$. In fact, what it is about to be proven, is that a Euclidean pseudo-dot product can be defined. This is necessary since many mathematical proofs on first and second order PMD rely on the possibility of performing such Euclidean dot products. The correct way of deducing this expression is to take into account (11), multiply it by the operator $\beta^j \breve{L}_j$ and obtain the trace of the product of generators' matrices (7):

$$\tau^k_z \breve{L}_k = \beta^k_\omega \breve{L}_k + \beta^i \tau^j \varepsilon_{ijk} \breve{L}_k \Rightarrow \tau^k_z \tau^l \breve{L}_k \breve{L}_l =$$
$$= \beta^k_\omega \tau^l \breve{L}_k \breve{L}_l + \beta^i \tau^j \tau^l \varepsilon_{ijk} \breve{L}_k \breve{L}_l \Rightarrow \tau^k_z \tau^l Tr[\breve{L}_k \breve{L}_l] =$$
$$= \beta^k_\omega \tau^l Tr[\breve{L}_k \breve{L}_l] + \beta^i \tau^j \tau^l \varepsilon_{ijk} Tr[\breve{L}_k \breve{L}_l] \Rightarrow$$
$$\tau^k_z \tau^l 2\delta_{kl} = \beta^k_\omega \tau^l 2\delta_{kl} + \beta^i \tau^j \tau^l \varepsilon_{ijk} 2\delta^k_l \Rightarrow \qquad (13)$$
$$\tau^k_z \tau_k = \beta^k_\omega \tau_k \Rightarrow \frac{1}{2}\frac{\partial (\tau^k \tau_k)}{\partial z} = \beta^k_\omega \tau_k \equiv$$
$$\equiv \frac{1}{2}\frac{\partial |\vec{\tau}|^2}{\partial z} = \vec{\beta}_\omega \cdot \vec{\tau}$$

One should bear in mind that no metric tensor whatsoever has been used in the present proof which is fully consistent with that nature of Poincare's method of representing the state of polarization of light.

The author would like to acknowledge the very helpful discussions and comments from M. Rodriguez Danta and C. Bellver Cebreros and regrets the fact that they did not want to sign this letter as co-authors.